\baselineskip=18pt
\def\a{\alpha}\def\c{\chi}
\def\f{\phi}
\def\l{\lambda}\def\m{\mu}\def\n{\nu}\def
\p{\pi}\def\r{\rho}\def\s{\sigma}
\def\y{\eta}\def\x{\xi}\def\z{\zeta}

\def\L{\Lambda}
\def\O{\Omega}

\def\na{\nabla}
\def\inf{\infty}\def\id{\equiv}\def\mo{{-1}}

\def\const{{\rm const}}

\def\mn{{\mu\nu}}

\def\af{asymptotically flat }\def\hd{higher derivative }
\def\fe{field equations }\def\bh{black hole }\def\as{asymptotically }
\def\coo{coordinates }

\def\ssy{spherically symmetric }
\def\dr{dimensionally reduced }\def\cp{critical points }

\def\hdim{higher dimensional }
\def\sch{Schwarzschild }\def\ads{anti-de Sitter }

\def\KK{Kaluza-Klein }\def\des{de Sitter }

\def\GB{Gauss-Bonnet }

\def\dsy{dynamical system }
\def\ab{asymptotic behavior }

\def\dsy{dynamical systems }

\def\section#1{\bigskip\noindent{\bf#1}\smallskip}

\def\PL#1{Phys.\ Lett.\ {\bf#1}}
\def\PRL#1{Phys.\ Rev.\ Lett.\ {\bf#1}}
\def\PR#1{Phys.\ Rev.\ {\bf#1}}\def\CQG#1{Class.\ Quantum Grav.\ {\bf#1}}
\def\NP#1{Nucl.\ Phys.\ {\bf#1}}
\def\JMP#1{J.\ Math.\ Phys.\ {\bf#1}}

\def\MPL#1{Mod.\ Phys.\ Lett.\ {\bf #1}} 
\def\PRep#1{Phys.\ Rep.\ {\bf#1}}

\def\ref#1{\medskip\everypar={\hangindent 2\parindent}#1}
\def\beginref{\begingroup
\bigskip
\centerline{\bf References}
\nobreak\noindent}
\def\endref{\par\endgroup}

\def\er{{\cal R}}\def\es{{\cal S}}\def\ef{e^{2\f}}
\def\eg{{\cal G}}
\def\qua{{(4)}}\def\np{{(n+4)}}

{\nopagenumbers \line{\hfil Revised version}
\line{\hfil 29 August 2006}
\vskip80pt
\centerline{\bf Black hole solutions of dimensionally reduced
Einstein-Gauss-Bonnet gravity}
\vskip40pt
\centerline{
{\bf S. Mignemi}\footnote{$^\ddagger$}{e-mail:
smignemi@unica.it}}
\vskip10pt
\centerline {Dipartimento di Matematica, Universit\`a di Cagliari}
\centerline{viale Merello 92, 09123 Cagliari, Italy}
\centerline{and INFN, Sezione di Cagliari}
\vskip100pt
\centerline{\bf Abstract}
\vskip10pt
{\noindent
We study the phase space of the \ssy solutions of the system obtained
from the dimensional reduction of the six-dimensional Einstein-\GB action.
We show the existence of solutions with nonflat asymptotic behavior.}
\vskip100pt\
P.A.C.S. Numbers: 97.60.Lf 11.25.Mj
\vfil\eject}

\section{1. Introduction.}

The possibility that higher-derivative corrections should be added
to the Einstein-Hilbert (EH) lagrangian of general relativity in order
to obtain a better behaved theory has been often considered.
Among the various possibilities, a preminent role is played by
the so-called Gauss-Bonnet (GB) terms.

GB lagrangians were introduced in [1] as the only possible generalization
of the EH lagrangian in higher dimensions that gives rise to \fe
which are second order in the metric, linear in the second derivatives,
and divergence free.
Another important property is that GB corrections do not introduce any new
propagating degrees of freedom in the spectrum of gravity [2].
However, since they vanish in lower dimensions unless nonminimally coupled to
scalar fields, they are mainly useful in the context of higher-dimensional
gravity, and especially Kaluza-Klein theories [3-5].
It must however be mentioned that GB contributions also appear in the low-energy
limit of string theories, [2,6], and may also play an important role in the
context of the braneworld scenario [7].

The introduction of GB terms in the action of \KK theories allows spontaneous
compactification of \hdim models without the need of introducing external fields.
For example, GB models admit ground states in the form of the
direct product of two maximally symmetric spaces [4]. They also have interesting
applications in higher-dimensional cosmology [5].

In order to explore further the physical implications of the dimensional reduction
of \hdim models of gravity including GB corrections, it is interesting to study
the existence of black
hole solutions of the dimensionally reduced theory, with compact internal space.
In the case of pure Einstein gravity, this investigation was performed in [8],
where it was shown that the only solution of physical interest is the
four-dimensional \sch metric with flat internal space.
In the GB case, one may expect the existence of a greater variety of solutions,
and in particular also black holes with \ads asymptotics.

Some black hole solutions of the Einstein-GB \fe are already known in different
physical situations, as spherical symmetry in higher dimensions [9] or GB-scalar
coupling in four dimensions [10,11].
In these cases it results a modification with respect to the Einstein case of the
short-distance behaviour of the solutions near the singularities,
but also asymptotic or global properties of the black hole may be altered.

In this paper, our aim is to classify all solutions of the Einstein-GB system
taking the form of a direct
product of a four-dimensional \ssy \bh with a maximally symmetric internal
space. Since a general discussion would be too involved, we shall limit ourselves
to the case of six dimensions, where the only relevant GB correction is quadratic in the
curvature and has the form $\es=\er^{\m\n\r\s}\er_{\m\n\r\s}-4\er^{\m\n}\er_{\m\n}+\er^2$.

A powerful technique for investigating this topic is the study of
the phase space of the solutions of the field equations. This
method has been used for example in the Einstein case [8].
As mentioned above, when a GB term
is added to the action, the field equations are still second order,
and linear in the second derivatives, but no longer quadratic
in the first derivatives.
This fact gives rise to several technical problems. In
particular, the potential of the dynamical system is no longer
polynomial, but presents poles for some values of the variables [11].

The result of our investigation is that physically relevant \bh solutions
exist that are not asymptotically flat.

Let us start by considering the $\np$--dimensional action
$$I=\int\sqrt{-g}\ d^\np x\ (\er^\np+\a\es^\np),\eqno(1.1)$$
where $\er^\np$ is the curvature scalar and $\es^\np$ the quadratic GB
term of the manifold and $\a$ is a coupling constant of dimension $[L]^2$.

We want to perform a dimensional reduction which casts the metric
in the form of a direct product of a
four-dimensional manifold with an $n$-dimensional space of constant curvature, whose
size is parametrized by a scalar field $\f$.
In general, contrary to the Einstein case, it is not possible to find an ansatz for
the metric of the Einstein-GB system that completely disentangles the scalar field $\f$
from the curvature in the dimensionally reduced action, except when the internal space
is flat. Therefore we maintain the usual ansatz
$$ds_\np^2=e^{-n\f}ds_\qua^2+e^{2\f}g_{ab}^{(n)}dx^adx^b,\eqno(1.2)$$
where $ds_\qua^2$ is the line element of the four-dimensional spacetime  and $g_{ab}^{(n)}$
is the metric of the $n$-dimensional maximally symmetric internal space, with
$\er_{ab}^{(n)}=\l_ig_{ab}^{(n)}$.
The action is dimensionally reduced to
$$\eqalignno{&I=\int\sqrt{-g}\ d^4x\Big[(1+2\a\l_i\,e^{-2\f})\er^\qua+\a\,e^{n\f}\es^\qua+
4n\a\,e^{n\f}\eg_\mn^\qua\na^\m\phi\na^\n\phi&\cr
&+\left({n(n+2)\over2}-(n^2-2n-12)\a\l_i\,e^{-2\f}\right)(\na\f)^2-{n(n+2)(n^2+n-3)\over3}
\ \a\, e^{n\f}(\na\f)^4&\cr
&+\l_i\,e^{-(n+2)\f}+(n-2)(n-3)\a\l_i^2e^{-\np\f}\Big].&(1.3)}$$

The \dr action contains the Einstein and the GB terms nonminimally coupled to a scalar field,
and a standard kinetic term and a potential for the scalar field. In addition, one has a
non-standard quartic correction to the kinetic term and a coupling between the Einstein
tensor $\eg_\mn$ and derivatives of the scalar field. Of course, both these terms yield second
order field equations. The action (1.3) displays some similarity with the string effective
action studied in [11], but contains additional terms.

In the following discussion, it is important to fix the possible ground states for the model.
These are taken to be the direct product of a four dimensional and an $n$-dimensional maximally
symmetric space, i.e.\ $\er^\qua_{\m\n\r\s}=\L_e(g^\qua_{\m\r}g^\qua_{\n\s}-g^\qua_{\m\s}
g^\qua_{\n\r})$,
$\er^{(n)}_{\m\n\r\s}=\L_i(g^{(n)}_{\m\r}g^{(n)}_{\n\s}-g^{(n)}_{\m\s}g^{(n)}_{\n\r})$.
Substituting this ansatz into the \fe derived from (1.1), one obtains
$$\eqalignno{&\a[(n-1)(n-2)(n-3)(n-4)\L_i^2+24\L_e^2+2n(n-1)\L_e\L_i]+(n-1)(n-2)\L_i+12\L_e=0,&\cr
&\a[n(n-1)(n-2)(n-3)\L_i^2+12n(n-1)\L_i\L_e]+n(n-1)\L_i+6\L_e=0.&(1.4)}$$
The system always admits the solution $\L_e=\L_i=0$, as in the Einstein case, but one can also
obtain solutions with nonvanishing curvature, namely de Sitter or anti-de Sitter.\footnote{$^*$}
{We are only interested in \bh with asymptotic regions, so we shall not consider the \des case further.}
Consequently, \bh solutions of (1.1) may have anti-de Sitter behavior at spatial infinity.
In the following we shall concentrate on the case $n=2$. Eq.\ (1.4) then admits a solution
$\L_e=-{1\over2\a}$, $\L_i=-{3\over10\a}$, i.e.\ $AdS\times H^2$ for $\a>0$, or $dS\times S^2$
for $\a<0$.

\bigbreak
\section{2. The dynamical system.}

Let us  consider the case $n=2$.
For the four-dimensional metric we adopt the \ssy ansatz [8]
$$ds_\qua^2=-e^{2\n}dt^2+\s^{-2}e^{4\z-2\n}d\x^2+e^{2\z-2\n}g_{ij}dx^idx^j,\eqno(2.1)$$
where $\n$, $\z$ and $\s$ as well as $\f$ are functions of $\x$ and $g_{ab}$ is the metric
of a two-dimensional maximally symmetric space, with $\er_{ij}=\l_eg_{ij}$. Of course,
in the case of physical interest, $\l_e>0$.

It is then convenient to define new variables
$$\c=2\z-\n-\f,\qquad\y=2\z-\n-2\f\eqno(2.2)$$
Substituting the ansatz (1.2), (2.1) into the action, performing some
integrations by parts, and factoring out the internal space,
the action can be cast in the form
$$\eqalignno{
I=\, -8\p&\int d^4x\bigg\{\s\,\Big[6\c'^2+3\z'^2+3\y'^2-8\c'\z'-8\c'\y'+4\z'\y'\Big]
-{1\over\s}(\l_ee^{2\z}+\l_ie^{2\y})&\cr
&+4\a e^{-2\c}\Big[\s(\y'-\c')(4\z'+3\y'-5\c')\l_ee^{2\z}+
\s(\z'-\c')(3\z'+4\y'-5\c')\l_ie^{2\y}&\cr
&-\s^3(\z'-\c')(\y'-\c')(11\c'^2+4\z'^2+4\y'^2+7\z'\y'-13\c'\z'-13\c'\y')-\l_e\l_i\,
{e^{2(\z+\y)}\over\s}\, \Big]\bigg\}.&(2.3)}
$$

As usual, the action (2.3) does not contain derivatives of $\s$, which
acts therefore as a Lagrangian multiplier enforcing the Hamiltonian
constraint.
Another relevant property of (2.3) is that, in spite of the presence of the \hd GB term,
it contains only first derivatives of the fields, although up to the fourth power, and
therefore gives rise to second order field equations. A further
interesting property is that, due to our choice of variables,
the action is invariant under the interchange of $\z$ and $\y$.

One can now vary (2.3) and then write the \fe in first order form in terms of
the new variables,
$$W=\c',\quad X=\z',\quad Y=\y',\quad U= e^\c,\quad Z=\ e^{\z},\quad V=\ e^{\y},\eqno(2.4)$$
which satisfy
$$U'=WU,\qquad Z'=XZ,\qquad V'=YV.\eqno(2.5)$$

Varying with respect to $\s$ and then choosing the gauge $\s=1$, one obtains
the Hamiltonian constraint
$$
\eqalignno{E\id &\ P^2+\l_eZ^2+\l_iV^2&\cr
&+{4\a\over U^2}\left[\l_e\l_iZ^2V^2+\l_eZ^2(Y-W)A+\l_iV^2(X-W)B-
3(X-W)(Y-W)C^2\right]=0,&(2.6)}$$
where
$$P^2=6W^2+3X^2+3Y^2-8WX-8WY+4XY,\qquad C^2=11W^2+4X^2+4Y^2+7XY-13WX-13WY,$$
$$A=4X+3Y-5W,\qquad B=3X+4Y-5W.$$
Variation with respect to $\c$, $\z$ and $\y$ gives rise to the other \fe
$$\eqalignno{
&2X'+2Y'-3W'+\Big\{{2\a\over U^2}\big[\l_eZ^2(2X+4Y-5W)+\l_iV^2(4X+2Y-5W)+22W^3&\cr
&\quad-2X^3-2Y^3-36W^2X-36W^2Y-12X^2Y-12Y^2X+17WX^2+17WY^2+44XYW\big]\Big\}'&\cr
&\quad={2\a\over U^2}\big[-\l_e\l_iZ^2V^2+\l_eZ^2(Y-W)A+\l_iV^2(X-W)B-
(X-W)(Y-W)C^2\big],&(2.7)
\cr&&\cr
&X'+2Y'-2W'+\Big\{{4\a\over U^2}\big[\l_eZ^2(2X+2Y-3W)-(X-W)(10W^2+2X^2+5Y^2+6XY&\cr
&\quad-9WX-14WY-\l_iV^2)\big]\Big\}'=\l_eZ^2+{4\a\over U^2}\big[\l_iV^2(X-W)B-(X-W)(Y-W)C^2\big],
&(2.8)\cr&&\cr
&2X'+Y'-2W'+\Big\{{4\a\over U^2}\big[\l_iV^2(2X+2Y-3W)-(Y-W)(10W^2+5X^2+2Y^2+6XY&\cr
&\quad-14WX-9WY-\l_eZ^2)\big]\Big\}'=\l_iV^2+{4\a\over U^2}\big[\l_eZ^2(Y-W)A-(X-W)(Y-W)C^2\big].
&(2.9)}$$
\smallskip
\noindent{In the variables (2.4), the problem takes the form of a six-dimensional
dynamical system, subject to a constraint. Notice that the function $E$
defined in (2.6) is a constant of the motion of the system (2.5), (2.7)-(2.9), whose
value vanishes by virtue of the Hamiltonian constraint. Since the system is obviously symmetric
for $V\to-V$, $Z\to-Z$, $U\to-U$, we shall only consider positive values of these variables.}
\bigskip
\noindent{\it The Einstein limit}
\smallskip
In the Einstein limit, $\a=0$, one recovers the results of [8].
We summarize them in terms of the variables introduced above:
when $\a=0$, the dynamical system reduces to eqs. (2.5) and
$$2X'+2Y'-3W'=0,\qquad X'+2Y'-2W'=\l_eZ^2,\qquad 2X'+Y'-2W'=\l_iV^2,\eqno(2.10)$$
subject to the constraint
$$E=P^2+\l_eZ^2+\l_iV^2=0.\eqno(2.11)$$
The physical trajectories lie on the four-dimensional hyperplane $E=0$.
Moreover, the system is independent of the variable $U$, and one may restrict the analysis
to $U=0$. It is evident that  $2X+2Y-3W$ is a constant of the motion for the system (2.10)
and therefore one of the variables, say $W$,
could be eliminated, but we keep it for comparison with the GB case.

The critical points at finite distance correspond to the small radius limit of the
solutions. They lie on the surface $Z_0=V_0=P_0=0$,
but only points with $X_0=Y_0=W_0$ correspond to regular horizons, while the others give
rise to naked singularities. The
eigenvalues of the linearized equations around the critical points are
$0(3)$, $X_0$, $Y_0$, $W_0$.

Since we are interested in solutions with asymptotic regions, we are led to
study the phase space at infinity, which corresponds to the large radius limit of the
solutions. This can be investigated defining new variables
$$t={1\over W},\quad x={X\over W},\quad y={Y\over W},\quad u={U\over W},
\quad z={Z\over W},\quad v={V\over W}.\eqno(2.12)$$
In terms of these variables, the field equations at infinity are then
obtained for $t\to0$, and read
$$\eqalignno{&\dot t=-2(v^2+z^2)t,\qquad\dot x=z^2+2v^2-2(v^2+z^2)x\qquad\dot y=2z^2+v^2-2(v^2+z^2)y,
&\cr&\qquad\dot u=(1-2v^2-2z^2)u,\qquad\dot z=(x-2v^2-2z^2)z,\qquad\dot v=(y-2v^2-2z^2)v,&(2.13)}$$
where a dot denotes $t\,d/d\x$.

The critical points at infinity are found at $t_0=0$ and

a) $\l_iv_0^2=\l_ez_0^2=0$, $x=x_0$, $y=y_0$, with $3x_0^2+3y_0^2+4x_0y_0-8x_0-8y_0+6=0$.

b) $\l_iv_0^2=0$, $\l_ez_0^2=1/4$, $x_0=1/2$, $y_0=1$.

c) $\l_ez_0^2=0$, $\l_iv_0^2=1/4$, $x_0=1$, $y_0=1/2$.

d) $\l_iv_0^2=\l_ez_0^2=3/16$, $x_0=y_0=3/4$.

{\noindent Points a) are the endpoints of the hypersurface $V=Z=0$, points b)
of the hypersurface $V=0$, points c) of the hypersurface $Z=0$. Exact solutions with
endpoints b), c) and d) are discussed in the appendix.}

The eigenvalues of the linearized equations around the critical points and their degeneracy
are:

a)\quad $0({\it3})$, $1$, $x_0$, $y_0$.

b, c)\quad $-1$, $-{1\over2}\,({\it3})$, ${1\over2}\,({\it2})$.

d)\quad $-{3\over2}$, $-{3\over4}\,({\it2})$, ${1\over4}$, $-{3\pm\sqrt15\over8}$.

The asymptotic behavior of the solutions can be deduced from the location
of the critical points at infinity [8]. Excluding points a) that do not correspond to physical
trajectories, one has, in terms of a radial variable $r$:

\halign{#&\quad#&\quad#\hfil&\qquad#\hfil\cr

&b) &$ds^2\sim -dt^2+dr^2+r^2d\O^2_+$,&$\ef\sim\const.$\cr

&c) &$ds^2\sim -r^2dt^2+r^2dr^2+r^2d\O^2_0$,&$\ef\sim r^2.$\cr

&d) &$ds^2\sim -r\,dt^2+dr^2+r^2d\O^2_+$,&$\ef\sim r.$\cr}

{\noindent Here, we have denoted with $d\O^2_+$ the metric of a unitary 2-sphere,
and with $d\O^2_0$ that of a flat 2-plane.}
The solutions ending at points b) are asymptotically flat, while the others have more exotic
behavior.

The phase space portrait is the following: solutions with regular horizons start at
$Z_0=V_0=0,\ $ $X_0=Y_0=W_0=a$, for some value of the parameter $a$, and end at
points b) if $\l_i=0$, or d) if $\l_i>0$. These last solutions, however, decompactify
for $r\to\inf$, since $\ef$ diverges in such limit. Also cylindrical solutions with $\l_e=0$
exist, which end at points c).
\bigbreak
\section{3. The Gauss-Bonnet phase space}

As discussed in section 1, in the GB case eqs.\ (1.4) admit the ground state
solution $\L_e=-1/2\a$, $\L_i=-3/10\a$ in addition to flat space, and therefore black holes
with anti-de Sitter \ab may be expected if $\a>0$.
The phase space of the system can be studied by the same methods used in the Einstein
case. As usual in the presence of \GB terms, some care must be taken because of the
poles at $U=0$. The limit $U\to0$ must be therefore taken, when necessary, at the end of the
calculations.

Equations (2.7)-(2.9) must be solved for the variables $X'$, $Y'$ and $W'$
in order to put the system in its canonical form. One can then find the critical points
at finite distance by requiring the vanishing of the derivatives of the fields.
As in the Einstein case, they lie on the hypersurface $U_0=Z_0=V_0=0$. However,
in the GB system, the other variables must satisfy the constraint $W_0=X_0=Y_0$,
or $X_0={4\pm\sqrt5\over5}\,W_0$,
$Y_0={4\mp\sqrt5\over5}\,W_0$. Only the first case corresponds to regular horizons.
In that case the eigenvalues of the linearized equations near the critical points
are identical to those found in the Einstein limit.

The critical points at infinity are obtained by writing the \dsy in terms of the variables
(2.12) and requiring the vanishing of their derivatives as $t\to 0$.
We find the following points:
\medskip
\def\noah{\noalign{\hrule}}

\halign{\strut#&\vrule\hfil\quad#\hfil\quad&\vrule\hfil\quad#\hfil\quad&
\vrule\hfil\quad#\hfil\quad&\vrule\hfil\quad#\hfil\quad&\vrule\hfil\quad#\hfil\quad&
\vrule\hfil\quad#\hfil\quad\vrule\cr\noalign{\hrule}
&&$x_0$&$y_0$&$u_0^2/\a$&$\l_ez_0^2$&$\l_iv_0^2$\cr\noah
&a)&1&1&0&0&0\cr\noah
&b)&1/2&1&0&1/4&0\cr\noah
&c)&1&1/2&0&0&1/4\cr\noah
&e)&2/3&1&2/9&0&$-1/15$\cr\noah
&f)&1&2/3&2/9&$-1/15$&0\cr\noah
&g)&$1$&1&$2$&$-1$&$-1$\cr\noah
&h)&4/5&4/5&6/25&0&0\cr\noah
&i)&2/3&1&0&1/3&0\cr\noah
&l)&1&2/3&0&0&1/3\cr\noah
&m$_\pm$)&${4\pm\sqrt5\over5}$&${4\mp\sqrt5\over5}$&0&0&0\cr\noah
 }
\medskip

{\noindent In case a), the curve of the previous section reduces to a single point.
Furthermore, the points d) have disappeared, except in the limit $\a\to0$ (i.e.\ $t\to\inf$).
The points with $u_0=0$ are attained by taking the limit $u_0\to0$ at the end of the calculation.
It is also evident that points e)-h) exist only if $\a>0$.

It is interesting to notice that the location of the critical points at infinity is very similar to
that of the pure Einstein system with a cosmological constant, investigated in [12],
except for the presence of the new points i), l) and m$_\pm$).
It seems therefore that one of the the main ingredients in fixing the structure of the phase space
at infinity, and hence the \ab of the solutions, is the relative dimension of the terms in the action.
For a detailed discussion see [13].}

From the eigenvalues and the eigenvectors of the linearized equations, one can deduce
the nature of the trajectories attracted by the various critical points at infinity.
The eigenvalues of the linearized equations near the \cp and their degeneracy are listed in the
following table, together with the nature of the trajectories attracted, for $W>0$.
\medskip
\halign{\strut#&\vrule\hfil\quad#\hfil\quad&
&\vrule\hfil\quad#\hfil\quad&\vrule\hfil\quad#\hfil\quad\vrule\cr\noalign{\hrule}
&&Eigenvalues (with degeneracy)&Trajectories attracted\cr\noah
&a)&$0\,({\it3})$, $1\,({\it3})$&\cr\noah
&b)&$-{1\over2}\,({\it3})$, $-1$, ${1\over2}\,({\it2})$&$\l_e>0$, $\l_i=0$\cr\noah
&c)&$-{1\over2}\,({\it3})$, $-1$, ${1\over2}\,({\it2})$&$\l_e=0$, $\l_i>0$\cr\noah
&e)&$-1\,({\it2})$, $-2$, $-{1\over3}$, $-{1\pm\sqrt{11/3}\over2}$&any $\l_e$, $\l_i<0$\cr\noah
&f)&$-1\,({\it2})$, $-2$, $-{1\over3}$, $-{1\pm\sqrt{11/3}\over2}$&$\l_e<0$, any $\l_i$\cr\noah
&g)&$-1$, $-2\,({\it2})$, 1, $-{1\pm i\sqrt{5/3}\over2}$&$\l_e<0$, $\l_i<0$\cr\noah
&h)&$-1\,({\it3})$, $-2$, $-{1\over5}\,({\it2})$&any $\l_e$, $\l_i$\cr\noah
&i)&$-{2\over3}$, $-{1\over3}$, $-1$, ${1\over3}\,({\it3})$&$\l_e>0$, $\l_i=0$\cr\noah
&l)&$-{2\over3}$, $-{1\over3}$, $-1$, ${1\over3}\,({\it3})$&$\l_e=0$, $\l_i<0$\cr\noah
&m$_+$)&$-{2\over 3}\,({\it2})$, 0, ${1\over3}$, ${2\pm3\sqrt{5}\over15}$&$\l_e>0$, $\l_i=0$\cr\noah
&m$_-$)&$-{2\over 3}\,({\it2})$, 0, ${1\over3}$, ${2\pm3\sqrt{5}\over15}$&$\l_e=0$, $\l_i<0$\cr\noah}
\bigskip

{\noindent The points a) do not attract any trajectory from finite distance.}

The \cp b)-c) generalize those found in the Einstein case, and have
the same asymptotic behavior. For what concerns the
other points,

\smallskip
\halign{#&\quad#&\quad#\hfil&\qquad#\hfil\cr

&e) &$ds^2\sim -r^2dt^2+r^{-2}dr^2+r^2d\O^2_+$,&$\ef\sim\const.$\cr

&f) &$ds^2\sim -r^4dt^2+dr^2+r^2d\O^2_-$,&$\ef\sim r^2.$\cr

&g) &$ds^2\sim -r^2dt^2+r^{-2}dr^2+d\O^2_-$,&$\ef\sim\const.$\cr

&h) &$ds^2\sim -r^2dt^2+r^{-1}dr^2+r^2d\O^2_0$,&$\ef\sim r.$\cr

&i) &$ds^2\sim -r^2dt^2+dr^2+r^2d\O^2_+$,&$\ef\sim\const.$\cr

&l) &$ds^2\sim -r^4dt^2+r^2dr^2+r^2d\O^2_0$,&$\ef\sim r^2.$\cr

&m$_\pm$) &$ds^2\sim -r^{2\pm\sqrt5}dt^2+r^{-(2\pm\sqrt5)}dr^2+r^2d\O^2_0$,&$\ef\sim
r^{1\mp\sqrt5}.$\cr}

\smallskip
\noindent{Here $d\O^2_-$ denotes the metric of a two-dimensional space $H^2$
with constant negative curvature.}

Of particular interest are the solutions that end at the critical point e), which arise
for positive $\a$. These asymptote to the exact ground state solution $AdS^4\times H^2$,
cited previously, that in the present \coo takes the form
$$ds^2=-\left({r^2\over2\a}+1\right)dt^2+\left({r^2\over2\a}+1\right)^\mo dr^2+r^2d\O^2_+,
\qquad\ef={10\a\over3}.\eqno(3.1) $$
Also interesting is the solution g), that asymptotes the exact solution $AdS^2\times H^2
\times H^2$.
Its four-dimensional section is analogous to a Bertotti-Robinson metric. The other
solutions have less common behavior.

The phase space structure can be summarized as follows: solutions with regular horizon
start at the points $U=V=Z=0$, $X=Y=W$ and terminate at one of the critical points listed above,
depending on  the values of $\a$, $\l_e$ and $\l_i$.

From the \KK point of view, the relevant solutions are those with $\l_e>0$ and
$\ef$ asymptotically constant. These are the \af (Schwarzschild-like) solutions with flat internal
space b), the \as\ads solutions with negatively curved internal space e),
 and possibly the more exotic the solutions i) and m$_+$), again with flat internal space,
whose nature is however not very clear.
\bigbreak
\section{4. Conclusions}
Higher-dimensional models of gravity naturally admit Gauss-Bonnet terms in the lagrangian.
Our study has shown that the compactification of the simplest model admits \bh solutions
displaying a variety of asymptotic behaviors. In particular, besides \af and asymptotically
\ads black holes, also physically reasonable solutions with \ab of type i) are present.

It turns out that the phase space of the model is quite similar to that of pure Einstein
gravity with a cosmological term [12]. It would be interesting therefore to consider the
effect of adding a cosmological constant to our model. This topic is currently under study
[13].

\medskip
{\noindent\bf Acknowledgements}

I wish to thank Maurizio Melis for valuable comments.

\section{Appendix A}
In the pure Einstein case,
exact solutions corresponding to the vanishing of $\l_i$ or $\l_e$
where obtained in [8] for generic spacetime dimensions. In our six-dimensional setting,
the solutions take the following form:
for $\l_i=0$ one obtains of course the \sch metric with constant scalar field,
$$ds^2=-(1-2M/r)dt^2+(1-2M/r)^\mo dr^2+r^2d\O^2_+,\qquad\ef=\const.$$
For $\l_e=0$, one has instead a solution of the form
$$ds^2=-r^2(1-2M/r)dt^2+r^2(1-2M/r)^\mo dr^2+r^2d\O^2_0,\qquad\ef=r^2.$$

In our case, a family of exact solutions with \ab d) can also be obtained, if one makes
the ansatz $\y=\z$. In fact, in this case, the \fe reduce to
$$4\z''-3\c''=0,\qquad 3\z''-2\c''=e^{2\z},$$
subject to the constraint $6\c'^2+10\z'^2-16\c'\z'+2e^{2\z}=0$.
Integrating the first equation, one obtains $\c'=4(\z'-c)/3$, and hence
$e^\c=A\,e^{4(\z-c\x)/3}$ for constant $A$ and $c$. Substituting in the second
equation, one obtains
$$\z''=3e^{2\z},$$
which is solved by
$$e^\z={2ae^{a\x}\over\sqrt3(1-e^{2a\x})}\ .$$
Regular black hole solutions satisfying the constraint are obtained for $c=a/4$.
For $\y=\z$, the metric functions of (2.1) are related to our variables by
$$e^\n=e^{3\z-2\c},\qquad e^\f=e^{\c-\z}.$$

Using these relations with $A=1$, defining $r=\int e^{2\z}d\x$, $r_0=2a/3$, and substituting
in (2.1), one finally obtains
$$ds^2=-{r-r_0\over r^{1/3}}\ dt^2+{r^{1/3}\over r-r_0}\ dr^2+r^{4/3}
d\O^2,\qquad\ef=r^{2/3},$$
or, in different coordinates,
$$ds^2=-R\left(1-{r_0\over R^{3/2}}\right)dt^2+{9\over4}\left(1-{r_0\over R^{3/2}}
\right)^\mo dR^2+R^2d\O^2,\quad\ef=R.$$

A more familiar expression can be obtained by writing the metric in its six-dimensional form:
$$ds^2=-\left(1-{r_0\over\hat r^3}\right)dt^2+\left(1-{r_0\over\hat r^3}\right)^\mo d\hat r^2
+{\hat r^2\over3}\,(d\O_i^2+d\O_e^2).$$
This is a variant of the well known six-dimensional Tangherlini metric, where the 4-sphere is
replaced by the direct product $S^2\times S^2$.

\bigskip
\section{Appendix B}
It may be interesting to write down some special exact solutions of the Einstein-GB system corresponding to the possible asymptotic behavior associated with the different critical points at infinity.
The properties of these solutions are more transparent in their six-dimensional form, in the Schwarzschild-like gauge
$$ds^2=-e^{2\l}dt^2+e^{-2\l}dr^2+e^{2\r}d\O^2_e+e^{2\s}d\O^2_i.$$
In this gauge is evident the presence of a symmetry for the interchange of $\r$ and $\s$, that follows from the
specific compactification considered. This entails a duality between points b), c) and e), f).
\medskip
\halign{#&\quad#&\quad#\hfil&\qquad#\hfil\cr
&b) &$ds^2=-dt^2+dr^2+r^2d\O^2_++d\O^2_0$.\cr

&c) &$ds^2=-dt^2+dr^2+d\O^2_0+r^2d\O^2_+$.\cr

&e) &$ds^2=-\left({r^2\over2\a}+1\right)dt^2+\left({r^2\over2\a}+1\right)^\mo dr^2+r^2d\O^2_++{10\a\over3}\,d\O^2_-$,\qquad or\cr
&&$ds^2=-{r^2\over2\a}\,dt^2+{2\a\over r^2}\,dr^2+r^2d\O^2_0+{10\a\over3}\,d\O^2_-$.\cr

&f) &$ds^2=-\left({r^2\over2\a}+1\right)dt^2+\left({r^2\over2\a}+1\right)^\mo dr^2+{10\a\over3}\,d\O^2_-+r^2d\O^2_+$,\qquad or\cr
&&$ds^2=-{r^2\over2\a}\,dt^2+{2\a\over r^2}\,dr^2+{10\a\over3}\,d\O^2_-+r^2d\O^2_0$.\cr

&g) &$ds^2=-\left({r^2\over2\a}-m\right)dt^2+\left({r^2\over2\a}-m\right)^\mo dr^2+2\a\,d\O^2_-+2\a\,d\O^2_-\,.$\cr

&h) &$ds^2=-{r^2\over6\a}\,dt^2+{6\a\over r^2}\,dr^2+r^2d\O^2_0+r^2d\O^2_0.$\cr}
\noindent{The parameter $m$ is an arbitrary constant.}

We were unable to find exact solutions in the other cases.
\vfill\eject

\beginref
\ref [1] D. Lovelock, \JMP{12}, 498 (1971).
\ref [2] B. Zwiebach, \PL{B156}, 315 (1985);
B. Zumino, \PRep{137}, 109 (1986).
\ref [3] J. Madore, \PL{A110}, 289 (1985);
S. Mignemi, \MPL{A1}, 337 (1986);
\ref [4] F. M\"uller-Hoissen, \PL{B163}, 106 (1985);
\CQG{3}, L133 (1986).
\ref [5] J. Madore, \PL{A111}, 283 (1985);
D. Bailin, A. Love and D. Wong, \PL{B165}, 270 (1985).
\ref [6] D.J. Gross and J.H. Sloan, \NP{B291}, 41 (1987).
\ref [7] N. Deruelle and T. Dole\v zel, \PR{D62}, 103502 (2000);
C. Charmousis and J.F. Dufaux, \CQG{19}, 4671 (2002);
S.C. Davis, \PR{D67}, 024030 (2003);
J.P. Gregory and A. Padilla, \CQG{20}, 4221 (2003).
\ref [8] S. Mignemi and D.L. Wiltshire, \CQG{6}, 987 (1989).
\ref [9] D.G. Boulware and S. Deser, \PRL{55}, 2656 (1985);
J.T. Wheeler, \NP{B268}, 737 (1986);
D.L. Wiltshire, \PL{B169}, 36 (1986).
\ref [10] S. Mignemi and N.R. Stewart, \PR{D47}, 5259 (1993);
 P. Kanti,  N.E. Mavromatos, J. Rizos,  K. Tamvakis and
E. Winstanley, \PR{D54}, 5049 (1996);
T.  Torii,  H.  Yajima, and K. Maeda, \PR{D55}, 739 (1997);
S.O. Alexeyev and M.V. Pomazanov, \PR{D55}, 2110 (1997).
\ref [11] M. Melis and S. Mignemi, \CQG{22}, 3169 (2005); \PR{D73}, 083010 (2006).
\ref [12] D.L. Wiltshire, \PR{D44}, 110 (1991).
\ref [13] M. Melis and S. Mignemi, in preparation.

\endref
\end